\documentclass[final]{svjour3}
\usepackage{graphicx}
\usepackage{subcaption}
\usepackage[font={small}]{caption}
\usepackage{rotating}
\usepackage{amssymb}
\DeclareMathAlphabet{\mathcal}{OMS}{cmsy}{m}{n}
\usepackage{mathptmx}
\usepackage{cite}
\makeatletter
\journalname{Journal of Low Temperature Physics}
\begin{document}

\newcommand{\hdblarrow}{H\makebox[0.9ex][l]{$\downdownarrows$}-}
\title{A Superconducting Phase Shifter and Traveling Wave Kinetic Inductance Parametric Amplifier for W-Band Astronomy}

\author{G. Che$^1$ \and S. Gordon$^1$ \and P. Day$^2$ \and  C. Groppi$^1$ \and R. Jackson$^1$ \and H. Mani$^1$ \and P. Mauskopf$^1$ \and H. Surdi$^3$ \and G. Trichopoulos$^4$ \and M. Underhill$^1$}

\institute{$^1$School of Earth and Space Exploration, Arizona State University, Tempe, AZ 85287, USA\\
\email{gche2@asu.edu}\\
$^2$Jet Propulsion Laboratory, Pasadena, CA 91109, USA\\
$^3$Department of Electrical and Computer Engineering, University of California Davis, Davis, CA 95616, USA\\
$^4$School of Electrical, Computer, and Energy Engineering, Arizona State University, Tempe, AZ 85281, USA}

\maketitle

\begin{abstract}

The W-Band ($75-110\; \mathrm{GHz}$) sky contains a plethora of information about star formation, galaxy evolution and the cosmic microwave background. We have designed and fabricated a dual-purpose superconducting circuit to facilitate the next generation of astronomical observations in this regime by providing proof-of-concept for both a millimeter-wave low-loss phase shifter, which can operate as an on-chip Fourier transform spectrometer (FTS) and a traveling wave kinetic inductance parametric amplifier (TKIP).
%As a phase shifter, our device provides the means to measure the non-linear phase shift $\Delta\phi$ resulting from DC biasing a superconducting transmission line (STL).
%The circuit consists of two planar dual polarization planar radial probes that  couple signals from circular waveguide to four superconducting transmission lines that connect the two symmetric probes for each polarization. The transmission lines for each pair of probes are the same length but one has a bias tee at each end to allow DC current to be run through the signal line in order to change the relative path length in the two lines by modulating the kinetic inductance of the superconductor. 
Superconducting transmission lines have a propagation speed that depends on the inductance in the line which is a combination of geometric inductance and kinetic inductance in the superconductor. The kinetic inductance has a non-linear component with a characteristic current, $I_*$, and can be modulated by applying a DC current, changing the propagation speed and effective path length. Our test circuit is designed to measure the path length difference or phase shift, $\Delta \phi$, between two symmetric transmission lines when one line is biased with a DC current.
%Along with a superconductor's nonlinearity parameter $I_*$, which we measured to be $\approx 0.4\; \mathrm{mA}$ for a $10\; \mathrm{\mu m}$ wide and $20\; \mathrm{nm}$ thick NbTiN line, $\Delta\phi$ is a key parameter needed to optimize a W-Band traveling wave kinetic inductance parametric amplifier (KIP) that offers substantial bandwidth, dynamic range, and noise performance improvements over state-of-the-art transistor-based amplifiers. As an on-chip Fourier transform spectrometer (FTS), our device offers a dramatic reduction in size compared to its mechanical counterpart and eliminates the need for moving parts.
To provide a measurement of $\Delta\phi$, a key parameter for optimizing a high gain W-Band TKIP, and modulate signal path length in FTS operation, our $3.6 \times 2.5\; \mathrm{cm}$ chip employs a pair of $503\; \mathrm{mm}$ long NbTiN inverted microstrip lines coupled to circular waveguide ports through radial probes. For a line of width $3\; \mathrm{\mu m}$ and film thickness $20\; \mathrm{nm}$, we predict $\Delta\phi\approx1767\; \mathrm{rad}$ at $90\; \mathrm{GHz}$ when biased at close to $I_*$. We have fabricated a prototype with $200\; \mathrm{nm}$ thick Nb film and the same line length and width. The predicted phase shift for our prototype is $\Delta\phi\approx30\; \mathrm{rad}$ at $90\; \mathrm{GHz}$ when biased at close to $I_*$ for Nb.
%, which is approximately a factor of $\approx3.45$ greater than that for NbTiN.


\end{abstract}

\section{Introduction}
Many astronomical observations at long wavelengths use coherent amplification of weak signals from the sky to enable readout. The figures of merit for an amplifier are gain, bandwidth, dynamic range, and noise performance. An ideal amplifier produces high, uniform gain over the entire observation band while exhibiting both high dynamic range and quantum-limited noise performance. Wideband amplifiers are used as the first stage in radio astronomy receivers \cite{Weinreb2009,Pospieszalski2012,GoddardMilne2001} and as intermediate frequency (IF) amplifiers for mm-wave-THz heterodyne receivers. The use of low noise first stage amplifiers could increase the instantaneous bandwidth and simplify the design of higher frequncy instruments. High electron mobility transistor (HEMT) amplifiers achieve $>20\; \mathrm{dB}$ gain over the entire ALMA Band 3 ($84-116\; \mathrm{GHz}$), but their best noise temperature is $\sim 25\; \mathrm{K}$, which is 5-10 times above the quantum limit \cite{Cuadrado-Calle2017,Tang2017, Samoska2012} and is a significant contribution to the system noise. Replacing HEMTs with an amplifier that simultaneously achieves high gain across multi-octave instantaneous bandwidth and quantum-limited noise performance would significantly improve the sensitivity of ALMA and similar instruments.

The traveling wave kinetic inductance parametric amplifier (TKIP) is an emerging technology that offers both wide instantaneous bandwidth and quantum-limited noise performance. Parametric amplifiers produce gain through four wave or three wave mixing (FWM/TWM) during which a strong pump mixes with a weak signal through a non-linear medium. Fiber optic amplifiers, which exploit the Kerr effect of non-linear optical materials, represent a well-established amplifier technology in the telecommunications industry \cite{Hansryd2002,Tong2011}. TKIPs, which exploit the non-linear kinetic inductance of superconductors, provide an analogous amplifier technology for mm-wave applications. Realized in superconducting transmission lines (STLs), which are inherently wideband, TKIPs achieve a maximum gain that depends on two superconductor material properties: $I_{*}$, the characteristic current parameter that sets the scale for non-linearity and $\Delta\phi_{\mathrm{max}}$, the maximum non-linear phase shift that can be induced by applying DC bias to a STL \cite{Eom2012Supp}. \cite{Bockstiegel2014,Adamyan2016,Vissers2016,Chaudhuri2017} have investigated TKIPs operating in the $10\; \mathrm{GHz}$ range that achieve $\sim15\; \mathrm{dB}$ gain over $\sim4\; \mathrm{GHz}$ of bandwidth and noise temperature as low as $0.5\pm0.3\; \mathrm{K}$, which approaches the quantum limit.

We have developed a dual purpose superconducting circuit with an inverted microstrip geometry that provides proof-of-concept for two technologies: a W-Band TKIP and a DC current controlled differential phase shifter which can act as an on-chip Fourier transform spectrometer (FTS). As a phase shifter, our circuit also provides a measurement of $\Delta\phi_{\mathrm{max}}$, which combined with $I_{*}$ from previous experiments, provides the parameters necessary to design and optimize a high-gain W-Band TKIP. The STLs on our device are not dispersion-engineered, but will still produce quadratic gain, demonstrating parametric amplification due to FWM/TWM at W-Band frequencies.
%As an on-chip FTS, our circuit offers a dramatic size reduction compared to its mechanical counterpart and eliminates the need for moving parts.
Here we describe our circuit design, fabrication process, test setup, and phase shift and gain measurements.

\section{Principle and Design}
\subsection{Kinetic Inductance Parametric Amplification}

For $T\ll T_c$, the kinetic inductance per unit length of a STL is
\begin{equation}
\label{eq:kineticInductance}
\mathcal{L}_{k}\left(I\right) \simeq \mathcal{L}_{k,0}\left[1+\left(\frac{I}{I_*}\right)^2\right],
\end{equation}
where $\mathcal{L}_{k,0}$ is the intrinsic kinetic inductance per unit length, $I_*$ is the characteristic current, and $I$ is the bias current applied to the line. The quadratic term in Eq. (\ref{eq:kineticInductance}) represents the non-linearity through which FWM/TWM occurs to generate gain in a STL. Three tones are involved in this process: a strong pump ($f_p$), a weak signal ($f_s$), and generated idler ($f_i$). The strong pump mixes with a weak signal along the STL converting two pump photons into a signal photon and an idler photon with frequency $f_i=2f_p-f_s$, thus amplifying the weak signal by drawing power from the pump. Stronger pump tones result in more gain, but we are limited by $I_{*}$, which corresponds to the maximum pump power before the onset of dissipation. Following \cite{Eom2012Supp}, we measured $I_*\approx0.4\; \mathrm{mA}$ at a readout power of $-68\; \mathrm{dBm}$ for a low $Q$ NbTiN resonator with $10\; \mathrm{\mu m}$ line width and $20\; \mathrm{nm}$ film thickness by monitoring its fractional detuning at increasing readout powers. $I_*$ sets a limit on the non-linearity and thus a limit on the gain that can be produced by a STL of a given material and geometry. NbTiN is a particularly suitable superconductor for TKIP technology due to its large non-linearity and low microwave loss. 

In the absence of an applied current, a tone propagates down a STL with speed $v_{p,0}=1/\sqrt{(\mathcal{L}_{k,0}+\mathcal{L}_m)\mathcal{C}}$, where $\mathcal{L}_m$ and $\mathcal{C}$ are the line's geometric inductance and capacitance per unit length, respectively, and attains a phase shift equal to the line's unbiased path length $\phi_0=2\pi fl/v_{p,0}$, where $l$ is its physical length. Applying a DC bias $I\leq I_*$ to the line, the  propagation speed and biased path length become $v_{p}\left(I\right)=1/\sqrt{(\mathcal{L}_{k}\left(I\right)+\mathcal{L}_m)\mathcal{C}}$ and $\phi\left(I\right)=2\pi fl/v_p\left(I\right)$, respectively, resulting in an additional non-linear phase shift $\Delta\phi\left(I\right)=\phi\left(I\right)-\phi_0$ compared to that attained on the unbiased line. This non-linear phase shift causes dispersion between the pump, signal, and idler tones, resulting in a predicted signal gain of $G_{s}=1+\left(\Delta \phi\right)^2$.
%, which defines the quadratic gain regime of FWM.
Phase matching these tones through dispersion engineering \cite{Chaudhuri2015} enables
%us to access the
the exponential gain behavior of FWM/TWM, namely $G_{s}=\exp\left(2\Delta\phi\right)/4$. In both cases, we need to measure $\Delta\phi$ to determine the maximum achievable gain, which defines the requirement on a TKIP's STL length.

\begin{figure*}[t!]
\centering
\includegraphics[width=\linewidth, keepaspectratio]{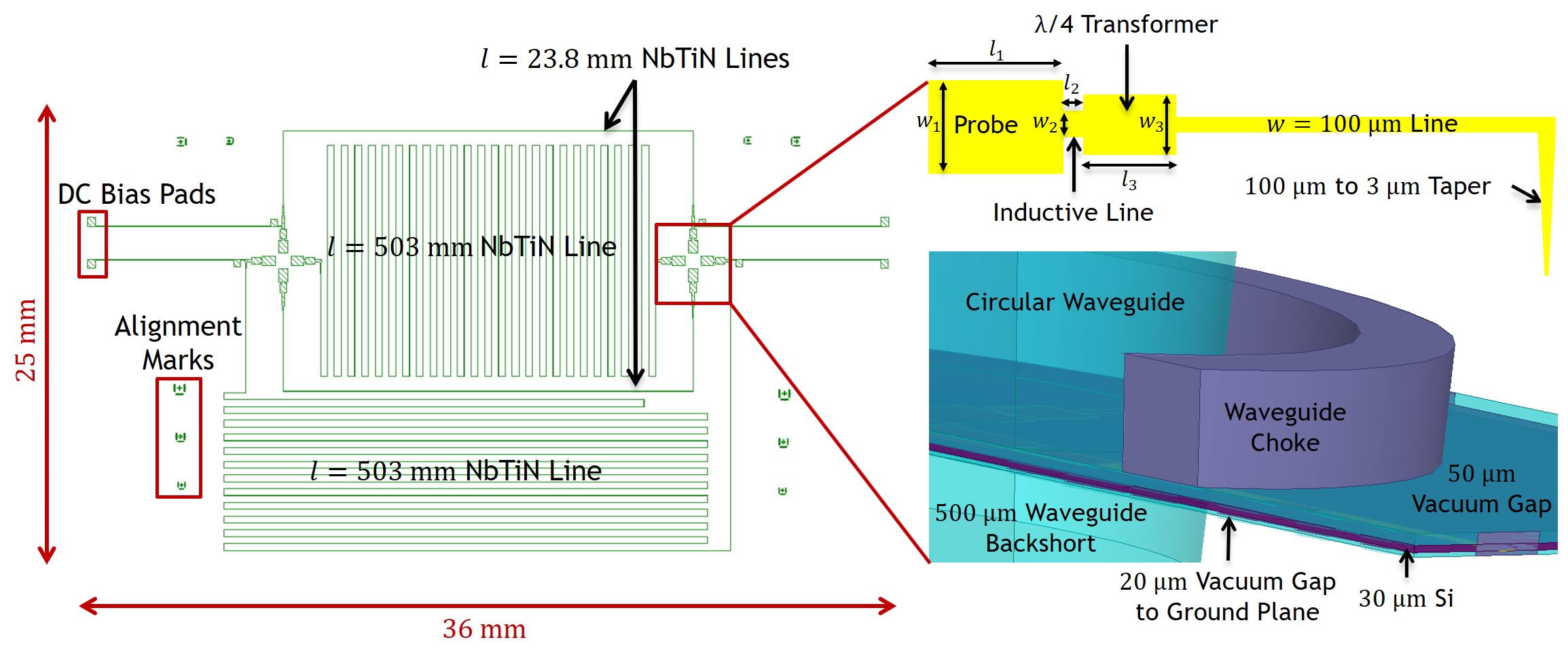}
\caption{W-Band phase shifter and TKIP circuit layout {\it (left)} with detailed top {\it (top right)} and cross-sectional {\it (bottom right)} views of the radial probe circular waveguide-to-inverted microstrip transition. Optimizing the dimensions for maximum coupling between the waveguide and inverted microstrip across the band yields the following: $w_1=0.4\; \mathrm{mm}$, $l_1=0.56\; \mathrm{mm}$, $w_2=0.1\; \mathrm{mm}$, $l_2=0.07\; \mathrm{mm}$, $w_3=0.3\; \mathrm{mm}$, and $l_3=0.4\; \mathrm{mm}$. (Color figure online.)}
\label{fig1:circuitDesign}
\end{figure*}

\begin{figure*}[t!]
    \centering
    \begin{subfigure}[t]{0.5\textwidth}
        \centering
        \includegraphics[scale=0.24]{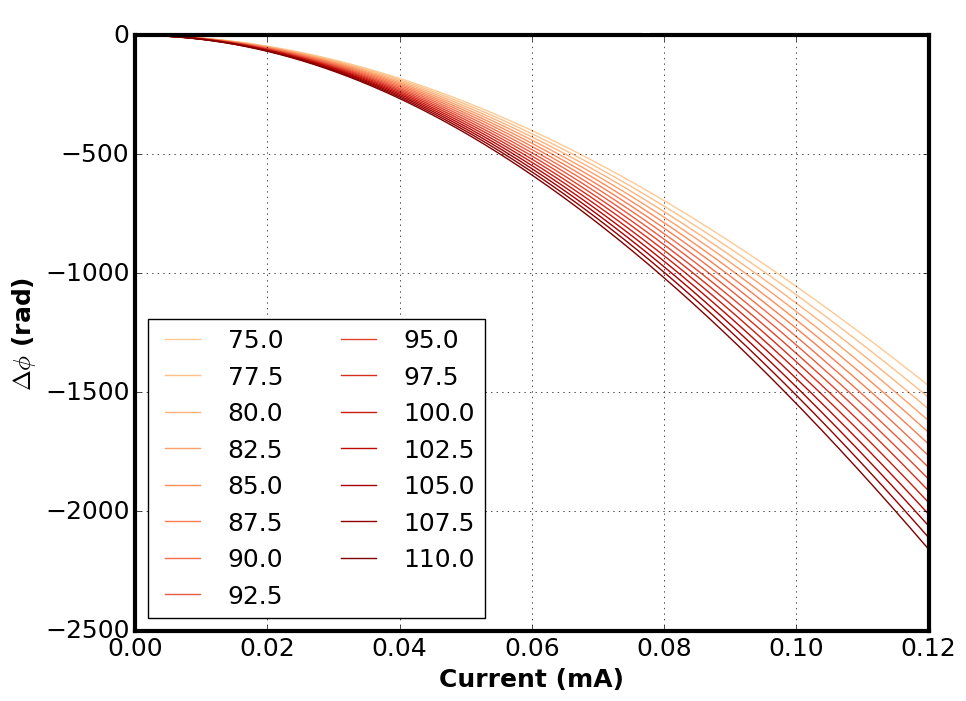}
        \caption{}
        \label{fig2a:quadraticGainPShip}
    \end{subfigure}%
    ~ 
    \begin{subfigure}[t]{0.5\textwidth}
        \centering
        \includegraphics[scale=0.25]{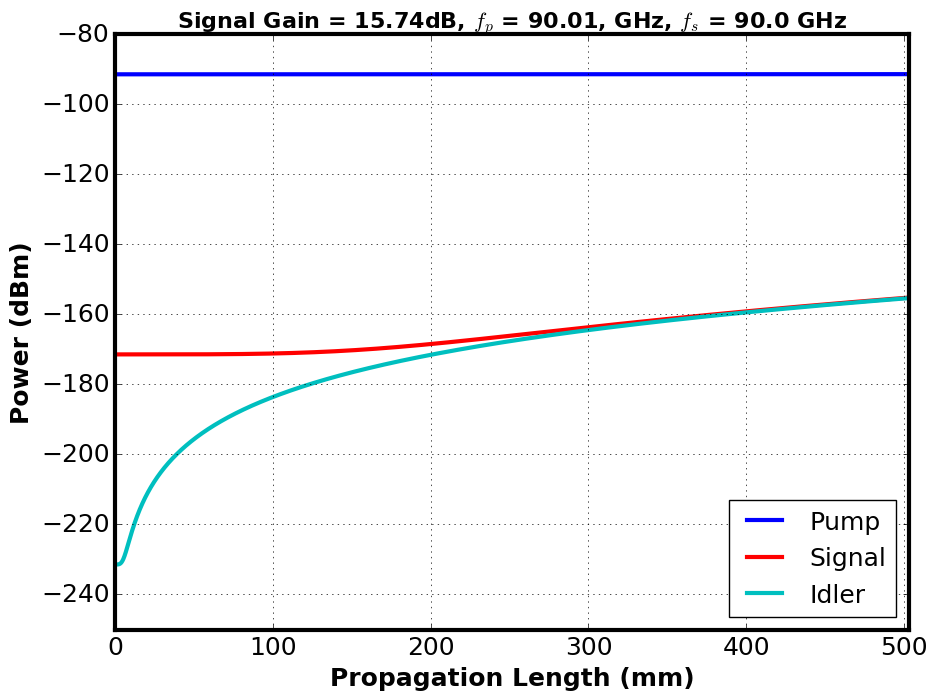}
        \caption{}
        \label{fig2b:exponentialGain}
    \end{subfigure}
    \caption{Predicted non-linear phase shift $\Delta\phi$ for our NbTiN circuit at various W-Band frequencies as a function bias current (a). Predicted signal gain due to FWM/TWM as a function of propagation length for $f_p\approx f_s=90\; \mathrm{GHz}$ (b). For a $-91\; \mathrm{dBm}$ pump, which is initially $80\; \mathrm{dB}$ above the signal, our line geometry produces $15.74\; \mathrm{dB}$ of signal gain. (Color figure online.)}
    \label{fig2:PredictedResults}
\end{figure*}

\subsection{Circuit Design}

To measure $\Delta\phi$, we developed the circuit shown in Fig.~\ref{fig1:circuitDesign}. The device is comprised of two pairs of $3\; \mathrm{\mu m}$ wide inverted microstrip lines with $20\; \mathrm{nm}$ NbTiN film thickness. %These dimensions are optimized with the relevant physics for FWM/TWM in W-Band.
A pair of identical $503\: \mathrm{ mm}$ long lines are used to measure $\Delta\phi$ and a pair of $23.8\: \mathrm{mm}$ long lines, which support a linear polarization orthogonal to that of the first pair, are used to calibrate the measurement setup. To measure $\Delta \phi$, one of the $503\: \mathrm{\mu m}$ lines is biased at $I\leq I_{*}$ with the other line unbiased. Radial probes couple a single frequency tone from an input waveguide port to both $503\; \mathrm{mm}$ lines. When they recombine at the output port, the signal on the biased line has been shifted an additional $\Delta\phi$ in phase relative to the signal on the unbiased line resulting in constructive or destructive interference. To determine $\Delta\phi$, we measure the complex transmission coefficient, $S_{21}$, at the output port as a function of bias current. Our choice of circular waveguide input and output allows us to access both microstrip polarizations without disassembling the setup.

Fig.~\ref{fig1:circuitDesign} also provides top and cross sectional closeup views of our radial probe circular waveguide-to-inverted microstrip transition (vice versa for the output port), which is based on designs from \cite{Shih1988,Fan1995,Leong1999,Datta2014}. Our design consists of a rectangular probe that intercepts the $\mathrm{TE}_{11}$ waveguide mode followed by a broadband impedance matching section. In the impedance matching section, an inductive line tunes out the probe's capacitance, a quarterwave transformer matches $2.46\; \mathrm{mm}$ diameter circular waveguide to $100\; \mathrm{\mu m}$ wide inverted microstrip across W-Band, and a taper transitions the line width from $100\; \mathrm{\mu m}$ to $3\; \mathrm{\mu m}$. We used HFSS and its MATLAB API to optimize all dimensions to achieve maximum coupling between the waveguide and inverted microstrip across the band. Beginning at the waveguide, the dielectric stack for our back-illuminated design consists of a $50\; \mathrm{\mu m}$ vacuum layer, $30\; \mathrm{\mu m}$ Si substrate, and $20\; \mathrm{\mu m}$ vacuum layer between the lines and ground plane, forming our inverted microstrip geometry. The waveguide itself is surrounded by a choke that attenuates higher order modes and terminated in a standard quarterwave backshort.

The normal state resistivity of NbTiN film has been measured to be $\rho_n\approx140\; \mathrm{\mu\Omega\,cm}$ \cite{Westig2013}. For $T\ll T_c$, the penetration depth is $\lambda_0=\sqrt{\hbar\rho_n/\pi\mu_0\Delta_0}$, where $\mu_0$ is the permeability of free space and $\Delta_0$ is the superconductor energy gap. With $T_c=14.5\; \mathrm{K}$ for NbTiN \cite{Westig2013}, $\lambda_0\approx325\; \mathrm{nm}$, so the intrinsic kinetic inductance per unit length for our STL is $\mathcal{L}_{k,0}=\mu_0\lambda_0^2/wt\approx2.22\; \mathrm{\mu H/m}$, where $w$ and $t$ are its line width and film thickness, respectively. For our inverted microstrip geometry, $Z_0=106\; \mathrm{\Omega}$ and $v_{p,0}=0.2c$, so $\mathcal{L}_m=1.79\; \mathrm{\mu H/m}$ and $\mathcal{C}=0.16\; \mathrm{nF/m}$. Applying $I_*=0.12\; \mathrm{mA}$ DC bias\footnote{We scaled $I_*$ obtained from our NbTiN resonator detuning measurement to our $w=3\; \mathrm{\mu m}$ and $t=20\; \mathrm{nm}$ line geometry assuming uniform current density.} to our $503\; \mathrm{mm}$ long line, $\Delta\phi\approx1767\; \mathrm{rad}$ at $90\; \mathrm{GHz}$. Fig.~\ref{fig2a:quadraticGainPShip} shows the predicted non-linear phase shift as a function of bias current across W-Band. As shown in Fig.~\ref{fig2b:exponentialGain}, FWM/TWM over $503\; \mathrm{mm}$ of NbTiN inverted microstrip produces $15.74\; \mathrm{dB}$ of signal gain for $f_p\approx f_s=90\; \mathrm{GHz}$ and a pump that is initially eight orders of magnitude stronger than the signal. To facilitate rapid proof-of-concept testing, we have fabricated a prototype on $200\; \mathrm{nm}$ thick Nb film while keeping the same geometry. $\rho_n\approx59\; \mathrm{\mu \Omega\,cm}$ and $T_c=9.2\; \mathrm{K}$ for Nb \cite{Westig2013}, so the penetration depth is $\lambda_0\approx84\; \mathrm{nm}$, which agrees with \cite{Anlage1989}. We are no longer in the thin film regime, so we use the full expression in \cite{DoyleDissertation2008} to calculate the intrinsic kinetic inductance for Nb to be $\mathcal{L}_{k,0}=22.7\; \mathrm{nH/m}$, which is much smaller than that for NbTiN. Biasing the line at $I_*$ for Nb, which we estimate to be a factor of $\sim3.45$ greater than $I_*$ for NbTiN with the same geometry, we predict $\Delta\phi\approx30\; \mathrm{rad}$ at $90\; \mathrm{GHz}$, which is still measurable using our network analyzer (VNA) setup.

\section{Fabrication and Packaging}

\begin{figure*}[t!]
    \centering
    \begin{subfigure}[t]{0.5\textwidth}
        \centering
        \includegraphics[scale=0.11]{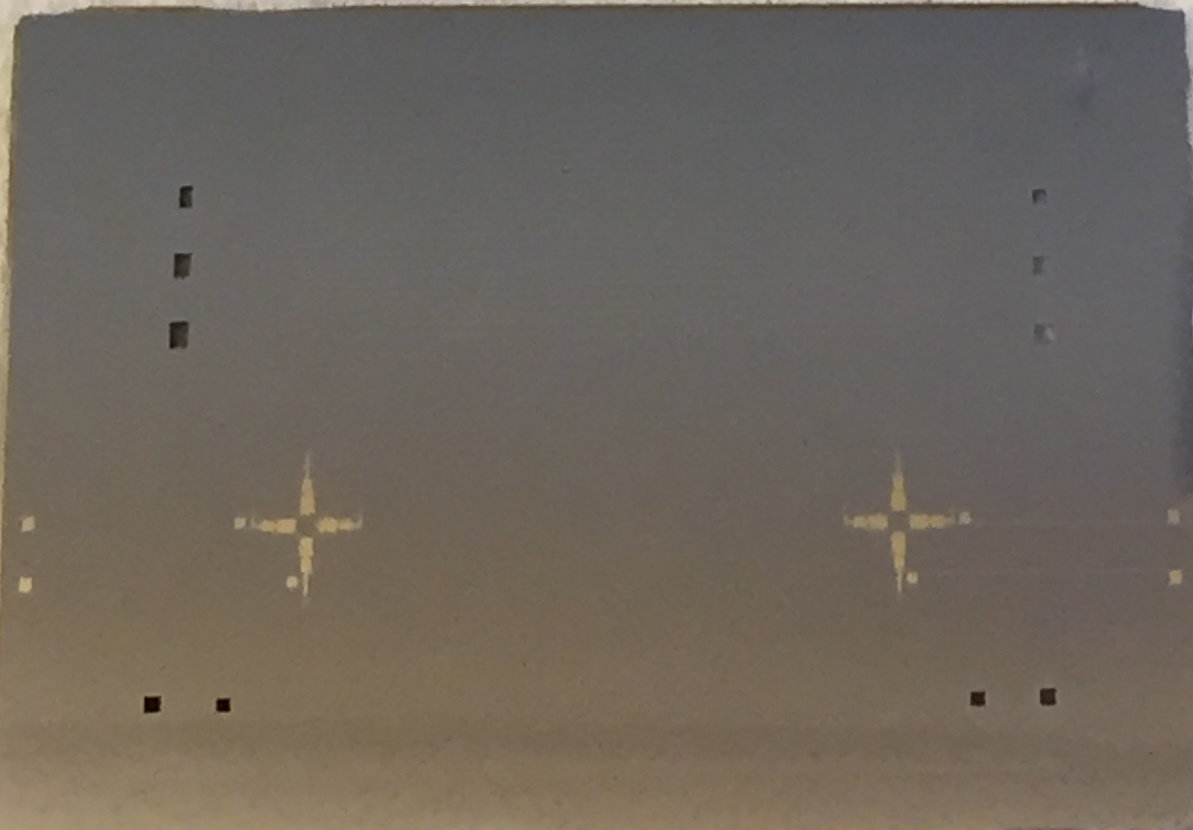}
        \caption{}
        \label{fig3a:deviceFabFront}
    \end{subfigure}%
    ~
    \begin{subfigure}[t]{0.5\textwidth}
        \centering
        \includegraphics[scale=0.15]{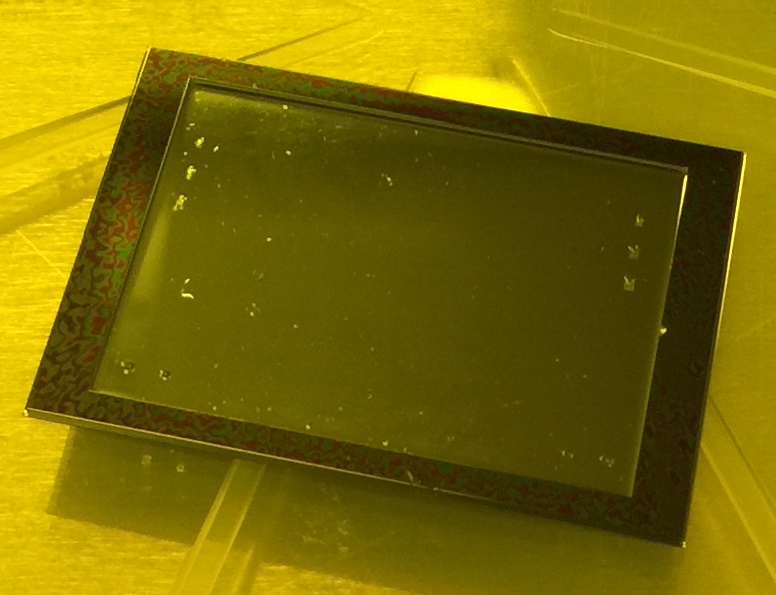}
        \caption{}
        \label{fig3b:deviceFabBack}
    \end{subfigure}%
    
    \begin{subfigure}[t]{\textwidth}
        \centering
        \includegraphics[scale=0.21]{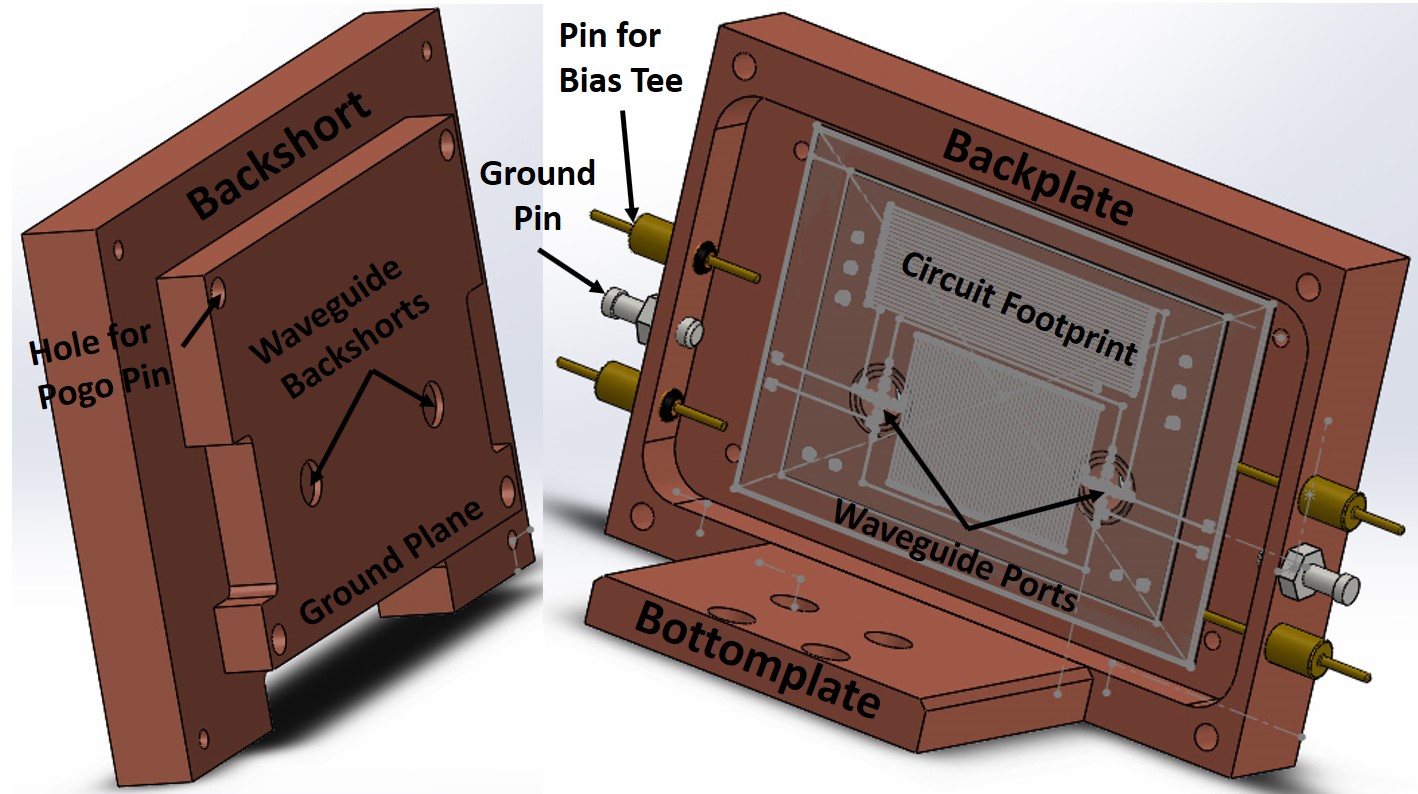}
        \caption{}
        \label{fig3c:packaging}
    \end{subfigure}
    \caption{Photos of the circuit side (a) and handle side (b) of our Nb prototype phase shifter and FTS circuit fabricated in the ASU NanoFab. On the handle side, a central rectangular feature is etched completely through both the handle Si and buried oxide layers, leaving only the $30\; \mathrm{\mu m}$ device layer across the circuit footprint. This feature maintains a $50\; \mathrm{\mu m}$ vacuum gap between the device Si and a central boss on our custom chip package's (c) Backplate piece, which also contains the waveguide ports and chokes. (Color figure online.)}
    \label{}
\end{figure*}

%Due to limited sputter deposition capability,
We have fabricated a prototype of our circuit using $200\; \mathrm{nm}$ thick Nb film in the ASU NanoFab. Beginning with a silicon-on-insulator (SOI) wafer comprised of a $30\; \mathrm{\mu m}$ high-resistivity device Si layer, $2\; \mathrm{\mu m}$ buried oxide layer, and $500\; \mathrm{\mu m}$ handle Si layer, we use DC sputtering to deposit $200\; \mathrm{nm}$ of Nb on the device layer. We then pattern our circuit on the device layer using contact lithography. Fig.~\ref{fig3a:deviceFabFront} shows the circuit side of a completed chip. After patterning the handle side, we etch these features completely through both the handle Si and buried oxide layers, simultaneously placing the circuit on a $30\:\mathrm{\mu m}$ Si membrane and separating the die from the wafer. A handle side view of a completed chip is provided in Fig. \ref{fig3b:deviceFabBack}.
To establish our inverted microstrip dielectric stack, provide waveguide ports, and implement DC bias capability, we have designed the three-piece copper chip package shown in Fig. ~\ref{fig3c:packaging}. The chip is mounted to the piece labeled Backplate, into which the waveguide ports, chokes, and holes for bias tee pins are directly machined. The waveguide backshorts are machined into the piece labeled Backshort, which also contains the ground plane. When Backplate and Backshort are fastened together, the former maintains a $50\:\mathrm{\mu m}$ gap between the waveguide port and bare Si while the latter maintains a $20\:\mathrm{\mu m}$ gap between the circuit and ground plane. Due to the precise dimensions required, this package will be manufactured in the ASU Micromachining Laboratory, which regularly achieves feature size tolerance of $\pm1\; \mathrm{\mu m}$. We will also sputter coat the inside of this package with $200\; \mathrm{nm}$ of Nb to avoid conductor loss. The third piece, Bottomplate, mounts to the $4\; \mathrm{K}$ stage of our cryostat.

\section{Measurement Setup}
To deliver W-Band signals to and from our device on the $4\; \mathrm{K}$ stage of our pulse-tube cooled cryostat, we have developed a custom waveguide feedthrough with thermal break and vacuum window designs based on \cite{Melhuish2016} and \cite{Ediss2005}, respectively. Eliminating the need to precisely align waveguide sections across a gap, our thermal break employs roughly-aligned conical horns separated by $2.5\; \mathrm{mm}$ and surrounded by baffling to absorb signal leakage from the gap. The window material we use is $0.5\; \mathrm{mil}$ thick mylar, which has $>95\%$ transmission across W-Band. After integrating this feedthrough into our cryogenic testbed, we will use W-Band VNA extenders to perform continuity measurements for both pairs of STLs on our Nb prototype circuit to calibrate transmission through the system. Using these extenders and bias tees mounted on our chip package, respectively, we deliver a $90\; \mathrm{GHz}$ tone to the feedthrough's input waveguide and sweep the DC current applied to one of the $503\; \mathrm{mm}$ long STLs from zero to $I_*$ to measure $\Delta\phi$ as a function of bias current. 

\section{Conclusions and Future Work}
We have designed a dual-purpose superconducting circuit to provide proof-of-concept for a W-Band TKIP and current-controlled differential phase shifter that can operate as an on-chip FTS, two key technologies for the next generation of W-Band astronomy. We have also fabricated a Nb prototype of this circuit and assembled a measurement setup to verify a predicted phase shift of $\Delta\phi\approx30\; \mathrm{rad}$ when one of a pair of symmetric STLs is biased near the critical current. Successful measurement of $\Delta\phi$ would verify one mode of operation and demonstrate the feasibility of fabricating future TKIP devices in the ASU NanoFab. We will then fabricate a NbTiN device, outsourcing film deposition, and use the same measurement setup to perform both phase shift and gain measurements, which would facilitate W-Band TKIP optimization and provide direct proof-of-concept for parametric amplification in this frequency band. Operation as an FTS will also be thoroughly explored.
%W-Band TKIP would be first of its kind.

\begin{acknowledgements}
This work is supported in part by NSF AST ATI grants 1407621 and 1509078. We also thank the exceptional staff members at the ASU NanoFab, especially Scott Ageno, Jerry Eller, and Kevin Nordquist, for their indispensable contribution to our device fabrication effort.
\end{acknowledgements}

\pagebreak
%\bibliographystyle{unsrt}
%\bibliography{Manuscripts-WBandPhaseShifter_2017_JLTP.bib}

\end{document}